\documentstyle[11pt,newpasp,twoside,epsf]{article}
\begin{document}

\title{{\it XMM-Newton} First Observation in the Pleiades}
\author{K. R. Briggs \& J. P. Pye}
\affil{X-ray Astronomy Group, Department of Physics and Astronomy,
University of Leicester, Leicester, LE1 7RH, UK} 

\begin{abstract}
We present the first results from a 40 ks Guaranteed Time
{\it XMM-Newton} pointing in the Pleiades. We detect almost all early--mid
dM members in the field and several very low mass (VLM) stars --
including the brown dwarf (BD) candidate Roque 9 -- and investigate
the variation of X-ray activity levels, hardness ratios and flare
frequency with spectral type down to the BD regime.
\end{abstract}

\vspace{-0.3cm}

\section{Introduction}
The Pleiades is key to our understanding of stellar activity and its
evolution, being the closest rich sample of stars at $\approx 100$ Myr
old. The core of the cluster was surveyed by {\it ROSAT}
PSPC (Stauffer et~al. 1994; Hodgkin, Jameson, \& Steele 1995:
HJS95; Gagne, Caillaut \& Stauffer 1995: GCS95; Micela et~al. 1996)
and HRI (Micela et~al. 1999: M99), studying the X-ray characteristics
of members down to the mid-M dwarfs ($M_{\rm bol}<10$). The
non-detection of fainter Pleiads suggested such fully-convective
stars, thought unable to harbour a solar-like magnetic dynamo, have
lower activity levels.

Our single pointing covers a smaller area, but {\it XMM-Newton}
(Jansen et~al. 2001) enables us to probe deeper, over a wider energy
range, with better angular and energy resolution than the PSPC, and
with continuous time coverage. In addition, recent infrared CCD
surveys (e.g. Pinfield et~al. 2000) have extended membership lists to
substellar objects. Here, we pursue the X-ray properties of Pleiads
into the fully-convective regime.

\section{Data Analysis and Membership Catalogue}
Our observation was centred on
the BD Teide 1 (03~47~18~+24~22~31 J2000) and performed on 1$^{st}$
September 2001 with the Thick filter in place in
each telescope. The data were processed using SAS v5.0 and filtered
for hot pixels and background flaring events, yielding 39.9 and 32.7
ks of good data in the EPIC PN and each MOS detector
respectively. Images were extracted with binsize 4$\arcsec$ in three energy
bands (A: 0.3--0.8; B: 0.8--1.5; C: 1.5--4.5 keV) for each EPIC
instrument. 

\begin{figure}[h]
\plotfiddle{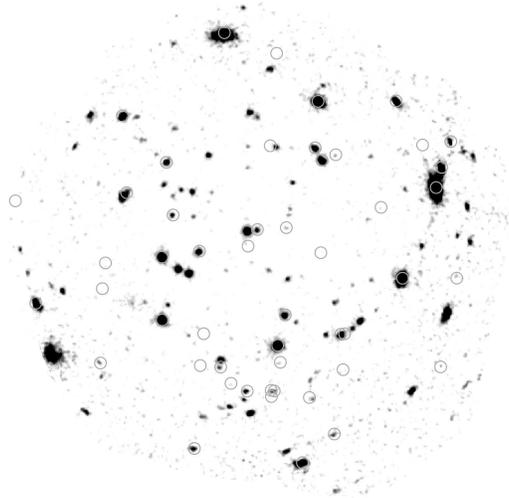}{6cm}{270}{50}{50}{-100}{200}
\caption{Mosaiced, smoothed, exposure-corrected, background-subtracted 
EPIC (PN + MOS1 + MOS2) image. Circles mark the positions of
catalogued members. The field is $\approx 15 \arcmin$ in radius.}
\label{fig_im}
\end{figure}

We constructed a list of members in the EPIC field of view from the
catalogues of Belikov et~al. (1998; B98) and Pinfield et~al. (2000;
P00), that are reckoned to be `complete' to $V=17.2$ and $I=19.2$
respectively (Fig.~\ref{fig_im}).
For each instrument, the triad of images was searched at the optical positions 
of these members using the Maximum Likelihood (ML) PSF-fitting routine
EMLDETECT (v4.2.2) with a ML threshold of 12 ($\cong 4.5 \sigma$). Simulations
yielded an estimate of $\approx 1$ false detection among our 51
members using this procedure. For the undetected Pleiads in each
image, $3 \sigma$ upper limits were estimated from the background
counts at the appropriate positions. Source counts were converted to
on-axis count rates before X-ray luminosities were estimated assuming a
distance of 127 pc, and conversion factors for each band consistent
with a `Mekal' plasma at $kT = 0.8$ keV with $N_{\rm H}=2\times
10^{20}$ cm$^{-2}$. 
Temperatures in the range 0.4--1.0 keV cause changes of $<7\%$ and
$<13\%$ in PN and MOS conversion factors respectively. Henceforth, all
luminosities are unabsorbed, in units of erg s$^{-1}$.

\begin{figure}[h]
\plotfiddle{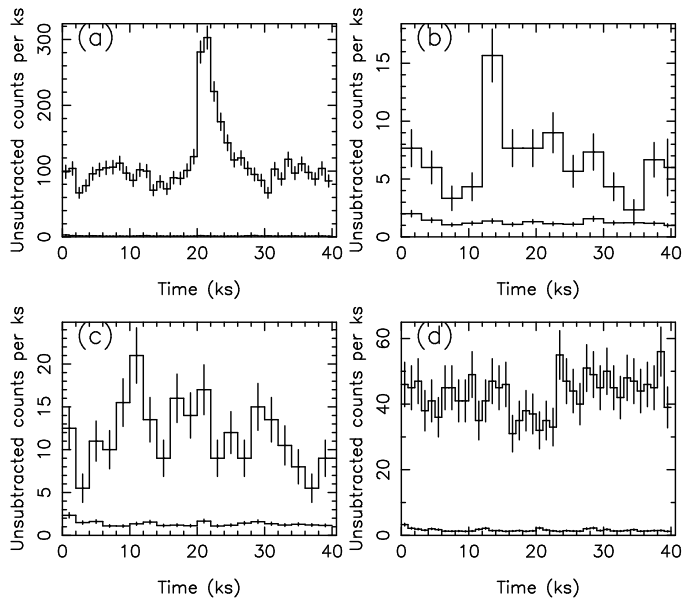}{9cm}{0}{50}{50}{-150}{-120}
\vspace{-1.2cm}
\caption{Sample lightcurves showing: (a) large flare on HII~1032; (b)
small flare on HHJ~252; (c) variability on HHJ~296; (d) `constancy' on
HII~1309. The upper (source) histogram in each plot is not
background-subtracted; the lower histogram shows the background lightcurve.}
\label{fig_lc}
\end{figure}

\section{Results and Discussion}
The PN detection threshold $log(L_{\rm X})$ varies
from $\approx 27.5$ on-axis to $\approx 27.7$ near the field edge. The
detection rates as a function of spectral type are summarised in
Table~\ref{tbl_det}. If we exclude four B98 dM stars discounted by P00, we 
detect almost all early--mid dM stars. We detect several late dM stars: 
in particular there is a PN B-band-only detection $\approx
10\arcsec$ from the BD candidate Roque 9. The chance of alignment with one of
the $\sim 50$ faint hard sources (probably background AGN) visible in
the X-ray image (Fig.~\ref{fig_im})
is $\sim 1/50$. The $log(L_{\rm X})$ of 27.7 is factor $>5$ more
than Neuhauser et~al.'s (1999) upper limit on Teide 1, and the
$log(L_{\rm X}/L_{\rm bol})$ of $-2.6$ is higher than the peak for
{\it Chandra}'s detection of a flare on the 500 Myr old BD LP 944-20
(Rutledge et~al. 2000). Roque 9 is poorly studied and the validity of
both its X-ray detection and its membership need confirmation.

PN lightcurves in the 0.3--4.5 keV range were extracted at the
position of each detected Pleiad using a radius of 20$\arcsec$ and a binsize
of 1 ks, and binned up to an average of $>15$ counts per bin. A
lightcurve was also extracted of the full, source-excluded field and scaled 
to the background level of each source position.
Flares were defined as peaking at factor $>$2 above the quiescent
level and non-flare-like variability was searched for using a $\chi^2$
test against constancy at the 95\% confidence level
(Fig.~\ref{fig_lc}). Large flares occurred on the bright sources
HII~1032 and HII~1100. Table~\ref{tbl_det} shows the variability and
flaring rates as a function of spectral type. K stars appear to have
the highest frequency of variability and flares, although we have
lower sensitivity to variability for the low count rates of the dM stars.

\begin{table}[h]
\caption{Detection and flare rates as a function of spectral type.}
\label{tbl_det}
\vspace{-0.4 cm}
\begin{center}
\scriptsize
\begin{tabular}{lccccc}
\tableline
Spectral type & B/A & F/G & K & M0-M5 & $>$ M5\\
\tableline
Mag. range & $6.5<V<8.3$ & $8.3<V<11.5$ & $11.5<V<14.7$ & $V>14.7$ \&
 & $I_{\rm C}>16.1$\\
 & & & & $I_{\rm C}<16.1$ & \\
Number in field & 5 & 3 & 5 & 19 & 15\\
Number confused & 0 & 0 & 0 & 2 & 2\\
Number detected & 3 & 3 & 5 & 16 & 5\\
Number variable & 1 & 1 & 3 & 8 & 0\\
Number flaring & 0 & 1 & 3 & 4 & 0\\
\tableline
\end{tabular}
\end{center}
\end{table}

\vspace{-0.3cm}

Fig.~\ref{fig_lx}a shows a general decrease of $log(L_{\rm X})$ with
absolute bolometric magnitude, $M_{\rm bol}$, mainly due to
smaller surface area, and hence available volume of X-ray-emitting 
plasma, with lower mass. Five flaring sources sit well above the
general trend. 
It is clear that the {\it ROSAT} PSPC threshold $log(L_{\rm X})$ of 28.5
prevented the detection of Pleiads with $M_{\rm bol} > 10$ and
lower-mass stars are capable of supporting hot coronae. There is a 
smaller scatter in $log(L_{\rm X})$ than seen in the {\it ROSAT}
results, but this is symptomatic of our smaller sample.

\begin{figure}[h]
\plottwo{}{}
\caption{Variation of (a) X-ray luminosity and (b) X-ray activity
level with bolometric magnitude. The symbols in (b) are as in (a).}
\label{fig_lx}
\end{figure}

Fig.~\ref{fig_lx}b demonstrates that although the X-ray luminosity
decreases to later types, the activity level as measured by
$log(L_{\rm X}/L_{\rm bol})$ increases. SHJ95 found $log(L_{\rm
X}/L_{\rm bol})$ continued increasing to $-2$ at the limit of their
survey, and the majority of {\it detected} low-mass stars were above
the canonical saturation level of $-3$, although a large number of upper 
limits caused the mean activity level to fall for $M_{\rm
bol} > 9$. Fig.~\ref{fig_lx}b shows our observation reaches the $-3$
level for all bar the faintest stars in our membership list and very few
objects stand above the saturation ceiling. The detections in the
range $10<M_{\rm bol}<11$ and upper limits for $M_{\rm bol}>11$
indicate the mean activity level turns over and falls toward later types.

\begin{figure}[h]
\plottwo{}{}
\caption{Variation of (a) HR1 (b) HR2 with bolometric magnitude.}
\label{fig_hr}
\end{figure}

Fig.~\ref{fig_hr} indicates the variation of coronal temperature with
$M_{\rm bol}$ through PN hardness ratios (HR): (a) HR1=$(B-A)/(A+B)$ and (b)
HR2=$(C-B)/(B+C)$, where $A$, $B$ and $C$ are the counts in the bands
A, B and C respectively. Gagne, Caillaut \& Stauffer (1995) found
coronal temperatures to be generally hotter for higher activity
levels. In their Chandra pointing, Krishnamurthi et~al (2001) saw a
slight tendency for lower-mass members to be harder. We find a suggestion 
that HR1 (a good tracer of temperature for a single-component Mekal
plasma in the range 0.2--1.0 keV) decreases toward later types while
HR2 (a good tracer of temperatures $>$ 1.5 keV) increases, explicable
if the coronal spectra of later types are flatter while those of
earlier types are strongly peaked in the 0.8--1.5 keV band,
although the effect is small and at the mercy of small number
statistics. Twelve of our Pleiads have sufficient counts
($>1000$) to perform EPIC PN spectroscopy.

\section{Summary and future work}
In a 40 ks {\it XMM-Newton} observation in the Pleiades, we:\\
(a) detect almost all members from F to M5, and several later than M5\\
(b) possibly detect the BD Roque 9: the X-ray detection must be
confirmed and optical spectroscopy is required to verify its membership\\
(c) find typical activity level turns over near $M_{\rm bol} \sim 10$: 
a wider, and deeper, X-ray survey is necessary to establish the mean
quiescent levels of VLM and substellar Pleiads\\
(d) see tentative evidence through hardness ratios that the coronae of 
dM stars have flatter spectra than those of A-G stars: twelve of 
the brightest sources have sufficient counts for EPIC X-ray
spectroscopy, and a deeper, wider survey is required to explore this further\\
(e) see variable behaviour across the full range of spectral
types, most frequently among K stars, although our sensitivity to
variability among dM stars is lower: large flares on HII~1032 and
HII~1100 deserve detailed analysis\\
(f) seek to investigate the effect of rotation rates -- an important
factor in coronal activity (e.g. M99) -- on this study.\\

\end{document}